\documentstyle[11pt,newpasp,twoside,epsf]{article}
\def\edcomment#1{\iffalse\marginpar{\raggedright\sl#1\/}\else\relax\fi}
\reversemarginpar

\begin{document}
\title{The Broad Emission Line Regions of Quasars: Current Status and Future Prospects}
\author{Gary Ferland}
\affil{Physics Department, University of Kentucky, Lexington, KY 40506}

\begin{abstract}
The spectrum emitted by the Broad Emission Line Regions of Active Galactic 
Nuclei can measure the luminosity of the central object and the chemical 
composition of the interstellar medium in a very young galaxy, and so 
constrain the expansion of the universe and the first stages of the 
evolution of massive galaxies. Here I review some recent developments in our 
understanding of the nature of the clouds and the interpretation of their 
spectra.

\end{abstract}

\section{Introduction}

The high redshift Active Galactic Nuclei (AGN) allow us to study the
spectrum of some of the earliest massive structures that formed. The
emission lines formed in the Broad Emission Line Region (BELR) tell us
of the matter that is most intimately associated with the innermost
regions.  The most fundamental questions we can ask of the BELR
spectrum include the following: What is the composition of the
emitting gas? What does it tell us about the evolution and age of the
associated star cluster? Can the spectrum be used as to deduce the
luminosity of a quasar, in turn allowing them to be used as
cosmological probes? What is the physics of the energy generation, and
what are the dynamics of the emitting gas? What is the mass of the
central object, and nature of the accretion mechanisms?

Although there are many questions, two things we know for sure are
that the emitting clouds are photoionized (line-continuum
reverberation measurements show this; see the review by Peterson
1993), and that the SDSS will produce the most complete database of
AGN spectra ever obtained. How can photoionization theory and the SDSS
database be combined to gain insight into fundamental questions about
AGN and the formation of the first massive structures in the redshift
5--10 universe?

Here I review what I consider to be the outstanding problems in quasar
emission-line research. This focuses on the problems that I have found
most interesting, and is not intended as a general review of AGN
emission lines.

\section{Physics of a Photoionized Cloud}

I described the physics of a single BELR cloud in Ferland (1999,
2003a). I only review the critical points here. The emitting gas has a
low density, by laboratory standards, and so is not in thermodynamic
equilibrium. As a result the observed spectrum is set by a host of
microphysical processes, and large-scale numerical simulations are
used to understand the message of the spectrum. The emitted spectrum
is sensitive to these details, which is a complication, but is also
why the lines reveal so much about the conditions within the gas.

Large-scale plasma simulations of the BELR were one of the first
applications of computers to astrophysics (Bahcall \& Kozslovsky 1969;
see also the excellent review by Davidson \& Netzer 1979). One
immediate conclusion was that the emitted spectrum was very sensitive
to cloud parameters. This led to the idea that some physical process
had ``fine-tuned'' these parameters, since the ensemble of quasars
have fairly similar emission-line spectra. This inhibited
emission-line research since the agent governing this unknown process
was itself unknown.

One popular model was the ``hot-warm'' pressure equilibrium picture of
Krolik, McKee, \& Tarter (1981). This assumed that the emission-line
clouds, assumed to be discrete structures with constant gas pressure
and temperatures near 10$^{4}$~K, were in pressure equilibrium with a
surrounding hot intercloud medium, presumed to be near the Compton
temperature of the continuum, $\sim10^{7}$~K. Although such
confinement could explain why a warm cloud with a special pressure
could survive, some problems were immediately noted. Mathews \&
Ferland (1987) pointed out that the hot intercloud medium would be
optically thick in the soft x-ray regime. Further, the two-phase
stability only occurs over a very limited range of gas densities (see
Figure 2 of that paper), while the theory does not explain why only
this narrow range of densities would be selected by most quasars.

\section{The Modern Revolution}

Two observations have dramatically revised our understanding of the
nature of the emission-line regions. The first was reverberation
mapping, which showed that the BELR was distributed over a broad range
of radii (Peterson 1993). The ``single parameter'' models described
above had placed the clouds at a single radius that was significantly
larger than observed.

The second discovery was that the BELR profiles remained smooth far
out in the line wings (Arav et al.\ 1998). The conventional picture
was that the BELR was composed of discrete ``clouds'', with each cloud
having only thermal motions within it, $\sim10~{\rm km~s}^{-1}$, for
H in a 10,000~K gas.  The observed line widths of $\sim10^{4}~{\rm
km~s}^{-1}$ were due to \textit{macro}turbulence, some sort of bulk
(perhaps orbital) motions of clouds, since non-thermal gas motion
would not occur within a particular constant gas pressure cloud. If
all clouds have the same emission-line spectrum then the observed
profiles reflect the number of clouds at each observed velocity
interval. Observations of the extreme line wings should detect single
clouds, which would be seen as single sharp components.

The first measurements by Capriotti, Foltz, \& Byard (1981) were
inconclusive, but Keck observations of Arav et al.\ (1998) show that
the line wings remain very smooth well into the extreme wings, to an
extent that essentially rules out discrete constant-gas-pressure
clouds.

\section{A Modern Picture}

In the remaining discussion the word ``cloud'' is used with the
broadest possible meaning. It could be a magnetically confined coronal
loop or a section of the outflowing wind above an accretion disk.

\subsection{The LOC Model}

We developed the ``locally optimally-emitting cloud'' (LOC) model of
the BELR to account for the distribution of cloud distances derived
from reverberation measurements (Baldwin et al.\ 1995). Gas exists over
a broad span of radii, and at each radius has a wide range of
densities. Selection effects, largely introduced by the atomic
physics, ensure that the ensemble of clouds produces the observed
spectrum (see also Baldwin 1997).

The most important conclusion of the LOC picture is that the observed
spectrum, being dominated by selection effects, is not determined by
detailed cloud properties or parameters. It is possible to avoid
questions of the cloud origins and move straight on to using the
spectrum to measure fundamental quasar properties that are of broad
interest. Some examples include the metallicity of the emitting gas,
what that says about the nuclear evolution of the central star
cluster, the shape of the energizing continuum that strikes clouds,
and the total mass in clouds. The key to using the BELR to measure
these properties is to first employ large-scale numerical plasma
simulations to identify the line ratios that are reliable diagnostics
of important quantities, despite the almost certain presence of
inhomogeneities and other complexities. This approach is discussed in
Ferland (2003b).

Some conclusions, not reviewed here, are that the metallicity of the
gas is at or above solar, even at redshift $z>5$, and that it
correlates with luminosity (Hamann \& Ferland 1999). This suggests
that the luminous quasar phase occurs only after stellar evolution has
proceeded long enough to increase the metallicity of the interstellar
medium to high values, and says something about the time scales for
formation of the deep potential well in which the quasar resides. The
mass associated with these clouds is substantial. Large reservoirs of
neutral or molecular gas, not emitting in the optical or UV, must
exist within the BELR if the properties of the emitting clouds are to
be stable -- for example a hydrogen ionization front must be free to
move across the cloud as the luminosity varies, so neutral gas must
exist. So a star cluster with a mass of order a galactic bulge is
responsible for the BELR if the gas is produced by normal stellar
evolution (Baldwin et al.\ 2003). For luminous quasars the total mass
in the BELR must be $\sim10^{4}$--$10^{5}\,M_{\odot}$.

\subsection{A Turbulent or Windy BELR}
\label{subsec:mylabel1}

The Arav et al.\ (1998) observation that line profiles do not break up
into distinct components rules out most forms of the discrete cloud
model. In a series of papers we investigated the possibility that
distinct clouds do exist, but that they are highly turbulent. This
added turbulence would make lines broader and so smear out the
spectrum.

Rees (1987) had suggested that BELR clouds may be magnetically
confined.  Non-dissipative magnetohydrodyamic (MHD) waves are usually
associated with a magnetic field in the ISM (Myers \& Goodman 1988)
and could provide an effective line broadening mechanism. In the
simplest form the magnetic energy density, $B^2/8\pi$, would equal
the kinetic energy in MHD waves, $\raise.5ex\hbox{$\scriptstyle
1$}\kern-.1em/ \kern-.15em\lower.25ex\hbox{$\scriptstyle 2$} \rho
u^2$, ($\rho$ is the gas density and $u$ the turbulent velocity) and
would be of order the gravitational energy density $GM\rho/r$. This
MHD turbulence would constitute \textit{micro}turbulence, a velocity
field that adds in quadrature with thermal velocities in affecting
line widths. In a series of papers Mark Bottorff and I investigated
the effects that such turbulence, or the fractal cloud distribution
that is argued to be an inevitable consequence of it, would have on
the observed spectrum (Bottorff \& Ferland 2000, 2001, 2002). The
conclusion was that the observed spectrum is consistent with very
large microturbulence velocities (Bottorff et al.\ 2000).

Another possibility is that the BELR is actually a wind, perhaps above
an accretion disk. Several models have been worked out, including
magnetically accelerated clouds (Bottorff, Korista, Shlosman, \&
Blandford 1997), models where the velocity gradient focuses line
escape (Chiang \& Murray 1996), and the wind models of K\"{o}nigl \&
Kartje (1994). This family of models also appears capable of
reproducing the observed spectrum.

The lesson in all this is that the physics that governs the spectrum
introduces the strong selection effects inherent in the LOC picture,
and so it is not too hard to reproduce the observed spectrum if clouds
are present with a wide variety of densities and separations from the
central object.  This is a disappointment --- the spectrum is not
strongly sensitive to the origin of the gas. However, it is also a
Good Thing --- we can skip over these uncertainties and go straight to
more fundamental questions about the quasar.

Although the detailed nature of the clouds remains a mystery, the
observations of smooth line profiles does show, conclusively, that
non-thermal motions must be present within BELR clouds. This has an
important implication for the cloud equation of state. When
non-thermal motions (microturbulence) are present the total pressure
(neglecting radiation pressure) is given by the sum of gas
($nkT$) and ram ($\raise.5ex\hbox{$\scriptstyle 1$}\kern-.1em/
\kern-.15em\lower.25ex\hbox{$\scriptstyle 2$} \rho u^2)$ pressures. If
the gas motions are supersonic, as required by the Arav et
al.\ observations, then ram pressure dominates over gas pressure by a
large amount. This conclusion, which follows directly from
observations, is important, since it rules out the entire family of
two-phase equilibrium models. We are left with either a
fractal/MHD/ISM-like or windy model for the regions.

\subsection{The Ionizing Continuum}

Mathews \& Ferland (1987; MF87) derived the mean shape of the ionizing
continuum of a typical quasar. We used direct observations where
possible, and the He~II $\lambda $1640 recombination line to measure
the unobservable continuum near 4 Ryd. This line, the $n$=3 to $n$=2
transition of He$^{+}$, is formed by Case B radiative recombination
even in the quasar environment.  This is because resonance lines of
He~II are destroyed before they undergo very many scatterings due to
absorption by H$^{0}$ and other species. As a result, substantial
populations of excited levels, the effect that complicates H~I lines
in this environment, do not occur. This essential simplicity was first
argued by MacAlpine (1981) and later confirmed by extensive
calculations (Bottorff et al.\ 2002). There are 0.57 $\lambda $1640
photons produced per He$^{+2}$ recombination. Since the number of
He$^{+2}$ recombinations per unit volume is equal to the number of
He$^{+}$ ionizations, the number of 4 Ryd photons striking gas is
1/0.57 = 1.7 times the number of $\lambda $ 1640 photons observed. The
$\lambda $1640 line directly measures the continuum near 200{\AA}.

The gas covering factor $\Omega $/4$\pi $ is the fraction of the
central object's ionizing photons that actually strikes emission-line
clouds, or, equivalent, the fraction of the sky as seen by the central
object that is covered by clouds. It must be significantly less than
unity since BELR clouds are never seen along the line of sight to the
continuum source (except possibly in BAL quasars).
Then, the total luminosity of the continuum source
at 4 Ryd is simply $\nu L_\nu \left( {4\,\mbox{Ryd}} \right) \approx
1.7L\left( {\lambda 1640} \right)\times \left( {\Omega / 4\pi }
\right)^{ - 1}$. Using this argument MF87 found a continuum shape that
peaked at an energy near 4 Ryd.

Zheng et al.\ (1997) directly measured the ionizing continuum in a
sample of high redshift quasars and found it to be surprisingly softer
than the MF87 continuum. The directly measured continuum shape is not
energetic enough to account for the observed high-energy lines when a
realistic covering factor is assumed (Korista et al.\ 1997). This is
evocative of the case in the Seyfert 2 galaxies where we do not
observe the same continuum as the emission line clouds (Antonucci
1993). So it seems likely that the ionizing continuum is beamed, and
that the continuum seen by BELR clouds is harder than what we observe
directly.

\section{Some Mysteries, and the Future}

What types of questions can a large data base like that produced by
SDSS answer? Here are some ideas.  But first an observation.

The idea that the BELR is emitted by distinct clouds, perhaps in
pressure equilibrium with a hot phase, has been around for some
time. As outlined above, this model has very serious problems, most
notably its inability to account for the smoothness of the emission
lines in the far wings. Arav's article in these proceedings reaches
similar conclusions from his study of broad absorption lines. Windy
and turbulent models were almost exclusively discussed at this
meeting. This represents a major shift in the cloud paradigm.

What might be done with a large data base such as the SDSS? The basic
questions that the emission lines can answer center on the
evolutionary state of the stellar and gas systems and their use as
cosmological probes.

The kinematic state of the BELR is still a mystery. Gaskell (1982) and
Wilkes \& Carswell (1982; see also Espey et al.\ 1989) discovered
that the low and high ionization emission lines do not have the same
redshift. Corbin (1992) investigated several implications of this. The
nature of this line shift is not now well understood. SDSS has already
investigated aspects of this (Richards et al.\ 2002). Does the extent
of this shift correlate with emission line relative intensities in any
way? A crucial but often overlooked constraint is that Ly$\alpha$ is
beamed back towards the continuum source in a photoionized cloud
(Ferland \& Netzer 1979). This means that we only see the far side
of the BELR in this line, unlike nearly all other lines, which are
radiated less anisotropically. The fact that the profiles agree to the
extent that they do is surprising.

If a quasar's luminosity could be determined from its spectrum then
they could be used as cosmological probes (see the conference
proceedings edited by Ferland \& Baldwin 1999). The Baldwin Effect
(BE; Baldwin 1977) is an inverse correlation between a line's
equivalent width (the strength of a line relative to the nearby
continuum) and a quasar's luminosity. This was discovered using the
CIV $\lambda $1549 line, although many lines show the
correlation. Korista, Baldwin, \& Ferland (1998) present a model of
the BE that is based on correlated changes between the continuum
shape, the gas metallicity, and luminosity.

Although line-to-continuum correlations like the BE have been
well-studied for some time, line to line correlations have been
elusive. This is the opposite from what is expected from
photoionization theory. The intrinsic emission-line spectrum of a
single cloud is set by the continuum shape, the metallicity, and the
selection effects described by the LOC. The relative intensities of a
pair of emission lines should be set entirely by these three
quantities. But the equivalent width of a line is affected by these
three things {\em plus} the gas covering factor and whatever
differences exist between the continuum we see at in the UV and the
ionizing continuum seen by the clouds. So the expectation is that
line-to-line correlations should be both simpler and cleaner than line
to continuum correlations.

It has been very hard to identify line to line correlations. Espey \&
Andreadis (1999) show that the slope of the BE depends on the
ionization potential of the species. The eigenvector analyses
described by Todd Boroson and Bev Wills in these proceedings detects
subtle changes in the spectrum.  These must be, at some level, a
correlation between relative emission line intensities. However, up to
now these hints at underlying correlations have been bedeviled by
insufficient data and selection effects.

With a complete enough data set the question could perhaps be inverted
--- what are the sources of scatter in the observed correlations and
could corrections to this scatter be introduced? Examples of possible
sources of noise include beaming of emission lines, chemical
inhomogeneities across the BELR, or if clouds only exist in portions
of the density --- incident continuum flux plane. Understanding the
scatter would be a step to making quasars valuable cosmological
probes.

These mysteries can be solved by going at this with the large SDSS
samples with their uniform and well-understood selection
procedures. For instance, photoionization theory predicts that the
spectrum is strongly affected by the shape of the ionizing
continuum. Does the line spectrum show any correlation with the slope
of the UV continuum, as expected by theory? Many avenues of research
are possible.

\acknowledgements

I thank the organizers of the conference for a wonderful gathering,
the NSF and NASA for support, and Jack Baldwin and Kirk Korista for
reviewing this paper.

\end{document}